\documentclass[MSNbibl,nameyear,dvips]{arxstspdf}
\usepackage{flushend}
\usepackage{stfloats}
\usepackage{graphicx}

\volume{27}
\issue{1}
\pubyear{2012}
\firstpage{3}
\lastpage{10}
\doi{10.1214/11-STS384}

\makeatletter

\newcommand{\tS}{\tilde{\Sigma}}
\newcommand{\tL}{\tilde{\Lambda}}
\newcommand{\tO}{\tilde{\Omega}}
\newcommand{\tQ}{\tilde{Q}}
\newcommand{\ZZ}{Z}
\newcommand{\sig}{\sigma}
\newcommand{\iid}{\mathop{\sim}^{\mathrm{i.i.d.}}}
\makeatother

\begin{document}
\begin{frontmatter}
\vspace*{12pt}
\title{Bayesian Nonparametric Shrinkage Applied to Cepheid Star Oscillations}
\runtitle{Bayesian Nonparametric Shrinkage}

\begin{aug}
\author[a]{\fnms{James} \snm{Berger}\corref{}\ead[label=e1]{berger@stat.duke.edu}},
\author[b]{\fnms{William H.} \snm{Jefferys}\ead[label=e2]{bill@astro.as.utexas.edu}}
\and
\author[c]{\fnms{Peter} \snm{M\"{u}ller}\ead[label=e3]{pmueller@math.utexas.edu}}
\runauthor{J. Berger, W. H. Jefferys and P. M\"{u}ller}

\affiliation{Duke University, University of Texas at Austin and
University of Vermont, and University of Texas at Austin}

\address[a]{James Berger is the Arts and Sciences Professor of
Statistics,
Department of Statistical Science,
Duke University, Durham, North Carolina 27708-0251, USA \printead{e1}.}
\address[b]{William H. Jefferys is Harlan J. Smith
Centennial Professor of Astronomy (Emeritus),
The University of Texas at Austin and
Adjunct Professor of Statistics,
The University of Vermont,
253 Strong Road,
Moretown, Vermont 05660, USA \printead{e2}.}
\address[c]{Peter M\"{u}ller is Professor,
Department of Mathematics,
The University of Texas at Austin,
1 University Station, C1200,
Austin, Texas 78712, USA
\printead{e3}.}

\end{aug}

%
\begin{abstract}
Bayesian nonparametric regression with dependent wavelets has
dual shrinkage properties: there is shrinkage through a dependent
prior put on functional differences, and shrinkage through the
setting of most of the wavelet coefficients to zero through Bayesian
variable selection methods. The methodology can deal with
unequally spaced data and is efficient because of the existence
of fast moves in model space for the MCMC computation.

The methodology is illustrated on the problem of modeling the
oscillations of Cepheid variable stars; these are a class of pulsating
variable stars with
the useful property that their periods of variability are strongly
correlated with their absolute luminosity. Once this relationship has
been calibrated, knowledge of the period gives knowledge of the
luminosity. This makes these stars useful as ``standard candles'' for
estimating distances in the universe.
\end{abstract}

%

\end{frontmatter}

\section{Introduction}

\subsection{Nonparametric Bayesian Shrinkage}

Bayesian analysis has long been a major methodological vehicle for
implementation of shrinkage ideas
in complex scenarios. There are two primary ways in which such
shrinkage is implemented. The
first is through use of prior distributions which shrink the unknowns
in some fashion---to prespecified locations or prespecified subspaces, depending
on the problem and type of prior. Thus an unknown normal mean could be
shrunk toward a specified prior mean;
a collection of unknown normal means could be shrunk toward the
hyperplane in which
the means are equal; and an unknown real function could be shrunk
toward the subspace of monotonic functions.
This is the Bayesian version of the type of shrinkage originating with
\citet{Stei56} and \citet{JaSt61}.

The second major Bayesian vehicle for shrinkage is Bayesian variable
selection, which sets some of the unknown parameters to zero. This is
often an overly drastic shrinkage, but is certainly not so in the
context of model selection, or in the context of
nonparametric function
estimation. In the latter setting, the unknown parameters that are set
to zero are typically coefficients of basis elements from a basis
representation of the function, and sparsity considerations strongly
encourage such shrinkage.

%
\begin{figure*}

\includegraphics{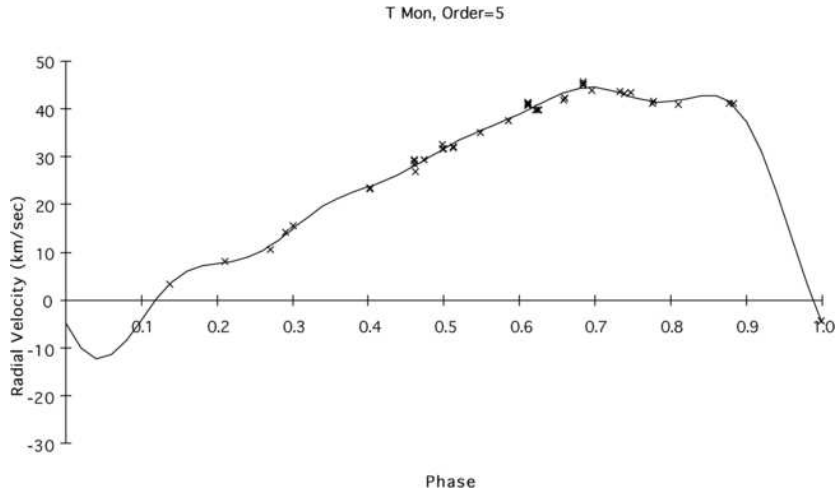}

\caption{The radial velocity data (the $\times$'s)
for T Mon, and their fit to a fifth-order trigonometric polynomial.}
\label{TMon}
\end{figure*}

Both of these shrinkage concepts are herein utilized in nonparametric
function estimation with
dependent wavelets. The motivating application is to Cepheid variable
stars and is described
in the next subsection; the functions to be estimated
can have arbitrary shapes, but are quite smooth. It is to induce
sufficient smoothness
that will utilize both ty\-pes of shrinkage discussed above.

\subsection{The Astronomical Problem}

There is a class of
stars, called Cepheid variables, that pulsate with a regular and
distinctive periodic signature.
The stars actually grow larger and then smaller, and as a result their
luminosities vary
periodically along with their colors. Since there is a~physical
relationship between
the star's linear diameter, its luminosity, and its color, there are
actually two independent
periodically varying quantities.

A very interesting and useful property of these stars is that their
mean luminosities
are highly correlated with their pulsation period, in that the
shorter-period stars are less luminous than the longer-period ones. This
is very well approximated as a linear relation between the log of the period
and the log of the luminosity. As a consequence, if one knows the slope and
intercept of this relationship, and measures the period of
a Cepheid (which is trivial), one can infer the luminosity with quite
high precision.
This makes these stars very useful as ``standard candles,'' because
knowledge of
a star's luminosity as well as its observed brightness allows us to compute
the distance from the inverse square law. Knowing the distance to the individual
Cepheid also gives us the distance to the galaxy or cluster of stars in which
it is embedded. Thus, these stars are fundamental in setting the
distance scale of the
universe.

The most challenging feature of the problem statistically is that the
key photometry and radial velocity curves
for a star are unknown, and have no simple structure.
In \citet{barnesal03}, Fourier polynomials of finite (but unknown)
degree were used to represent
these two curves. For instance, Figure \ref{TMon} presents the data
concerning the radial velocity of the surface
of the star T Moncerotis, at various phases of the star's period
(the actual data are indicated by the $\times$'s) together with a fifth-order
trigonometric polynomial fit
to the data.
Because of the possibility of quite arbitrary shapes for the photometry
and velocity curves for Cepheid variable stars,
we instead desired to model the curves via much more flexible wavelet
decompositions.

\subsection{Computational Implementation}
Posterior inference in this setup
is formally equivalent to variable selection in a normal linear
regression problem with massively many candidate covariates.
Posterior simulation requires averaging and/or selection across
alternative models defined by the set of basis functions (wavelets)
which are included in the model.
In the context of normal-linear regression, common approaches are
guided search in the model space using the Occam's Window principle
(\citep{madiganraftery94}; \citep{rafteryal97});
Markov chain Monte Carlo simulation across the model space
(\citep{georgemcculloch97}; \citep{smithkohn96});
and importance sampling or Gibbs sampling based on analytic
approximations to the marginal posterior distribution on the model
indicator
(\citep{clydedesiparm96};
\citep{clydeparmvida98}).
See, for example,
\citet{clyde99}, Hoe\-ting et~al. (\citeyear{hoetingal99})
and \citet{clydegeorge04}
for reviews.
In this paper we introduce a scheme for fast posterior simulation
across the model space, marginalizing over the wavelet coefficients.
We use a computational strategy similar to that used by
\citet{georgemcculloch97} and \citet{smithkohn96}
to allow fast computation of marginal model probabilities when
considering models differing by only one wavelet basis function.

\section{Wavelet Representation}
Wavelet decomposition allows representation of any square integrable
function $f(x)$ as
%
\begin{equation}
\quad f(x) = \sum_{k\in Z} c_{J_0k} \phi_{J_0k}(x) +
\sum_{j \geq J_0} \sum_{k\in Z} d_{jk} \psi_{jk}(x).
\label{eqdwt}
\end{equation}
Here $\psi_{jk}(x) = 2^{j/2} \psi(2^j x -k)$
and $\phi_{jk}(x) = 2^{j/2} \cdot \phi(2^j x -k)$
are wavelets and scaling functions at level of detail $j$ and shift
$k$.
In the context of statistical modeling, (\ref{eqdwt}) allows for
inference about random functions by defining a
probability model for the coefficients
$\theta=(c_{J_0k},d_{jk},$ $j \geq J_0;~k \in\ZZ)$,
that is, (\ref{eqdwt}) provides a parameterization of a random function
$f$ in terms of the wavelet coefficients $\theta$.
See, for example,
\citet{vidamuel99} or
Ferreira and Lee (\citeyear{ferreiralee07}), Chapter~5,
for a review of wavelet representations
relevant for statistical modeling.\looseness=1

Perhaps the most common application of (\ref{eqdwt}) in~%
sta\-tistical modeling is to nonlinear regression where~$f(x)$
represents the unknown mean response $E(y|x)$ for an observation $y$ with
covariate $x$.
\citet{chipal97},
\citet{clydeparmvida98},
\citet{vida98},
\citet{semadenial04},
\citet{mahletal05},
\citet{WangWood06},
\citet{terBraak06} and
Abra\-movich, Angelini and De~Canditiis (\citeyear{abramovichal07}),
among many others,
discuss Bayesian inference in such models assuming equally spaced data,
that is,
covariate values $x_i$ are on a regular grid.
For equally spaced data the discrete wavelet transformation is
orthogonal. Together with assuming independent measurement errors and
a priori independent wavelet coefficients this leads to
posterior independence of the $d_{jk}$. Thus the problem essentially
reduces to a sequence of univariate problems, one for each wavelet
coefficient.
See, for example,
\citet{yaukohn99}
for a review.
Generalizations of wavelet techniques to non-equidistant (NES)
design impose additional con\-ceptual and computational burdens.
A reasonable approximation is to bin observations in equally spaced
bins and proceed as in the equally spaced case. If only few
observations are missing to complete an equally spaced grid, treating
these few as missing data leads to efficient implementations
(An\-toniadis,~Gr{\'e}goire and McKeague (\citeyear{antoniadisal94}); \citep{caibrown97}).
We propose instead an approach\break which does not depend on
posterior independence. Our approach includes informative dependent
priors with positive prior probabilities for vanishing wavelet
coefficients.\looseness=1

\vspace*{2pt}
\section{\texorpdfstring{Shrinkage of \textit{\lowercase{f\textup{(}x\textup{)}}}}{Shrinkage of $f(x)$}}
\vspace*{2pt}

\subsection{Shrinkage Toward a Smooth Subspace}
\label{secsubspace}

Because of the wavelet representation that will be used, a function space
prior can be defined by considering the function at the discrete points
$\{i/n, i=1,\ldots,n\}$, where $n=2^J$. Letting $f_i = f(i/n)$, consider
the difference process $d_i = f_i-f_{i-1}$.

A function space prior that ``shrinks toward\break smoothness'' can be defined by
imposing positive~cor\-relations on the $d_i$.
Specifically, let
$d=(d_1,\ldots,d_n)$, and define the prior to be $p(d) = N(0,\Delta)$ with
$\Delta_{ij} = \lambda\exp(-\beta|i-j|)$; that is, we assume a
multivariate normal prior with scale parameter $\lambda$ and log
correlations proportional to distance.

%
\begin{figure*}

\includegraphics{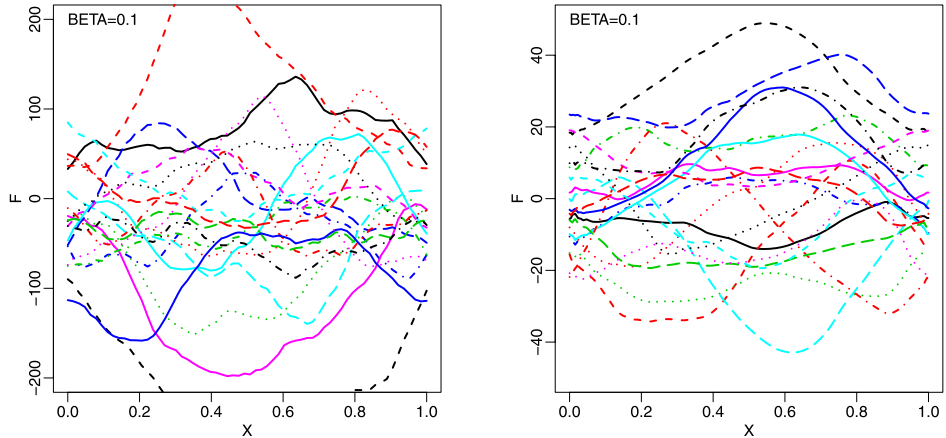}

\caption{For $\beta=0.1$, the left panel plots simulations from the prior
process on the
unknown function conditioning on \textit{all} wavelet coefficients
included; the right panel shows for comparison prior simulations
conditional on setting those coefficients equal to zero which are
excluded by the universal wavelet thresholding rule with $\sqrt
{2n}\hat
\sig$ of \textit{Donoho and Johnstone} (\protect\citeyear{donohoejohnstone94}).}
\label{figprior}
\end{figure*}

Let $\Delta_{(11)}$ denote the left upper $(n-1)\times(n-1)$ submatrix
of $\Delta$ and partition $\Delta$ into
\[
\Delta= \left[
\matrix{
\Delta_{(11)} & \Delta_{(12)} \vspace*{2pt}\cr
\Delta_{(21)} & \Delta_{(22)}}
\right].
\]
Let $v=\operatorname{Var}(\sum_{i=1}^n d_i) =
\lambda\sum_{i=1}^n \sum_{j=1}^n \exp(-\beta|i-j|)$.
Assuming $f_0 \sim N(0,\lambda\sigma^2_0)$ we find
\[
p(f_0,\ldots,f_{n-1} | f_0=f_n) = N(0,\lambda V) ,
\]
with $V = A H_0 A'$,
\begin{eqnarray*}
A &=& \left[
\matrix{
1 & 0 & \cdots& 0\vspace*{2pt}\cr
1 & 1 & \cdots& 0\vspace*{2pt}\cr
\ldots\vspace*{2pt}\cr
1 & 1 & \cdots& 1}
\right],\\
H_0 &= &\left[
\matrix{
\sigma^2_0 & 0 \vspace*{2pt}\cr
0 & H}
\right]
\quad\mbox{and}\\
H &=& \Delta_{(11)}-\Delta_{(12)} \Delta_{(12)}'/v .
\end{eqnarray*}

In view of the normalization property, $\Vert\phi_{jk}\Vert=1$, scaling
coefficients at the highest level of detail $J$ are approximately
proportional to the represented function,
$c_{Jk} \approx2^{-J/2} f_k$.
Therefore the multivariate normal prior on $(f_0,\ldots,f_{n-1})$
implies
$
p(c_J) = N(0,\allowbreak r_J\cdot\lambda V)
$
where $r_J = 2^{-J}$. Following common practice in the use of wavelet
decomposition, we will ignore\vadjust{\goodbreak} the proportionality constant $r_J$ and
assume
\[
p(c_J) = N(0,\lambda V).
\]
As long as we also drop $r_J$ in the reconstruction of $f(x)$, ignoring
the proportionality constant will leave the final inference unchanged.

The prior $p(c_J) = N(0,\lambda V)$ implies a dependent multivariate normal
prior for the vector of all wavelet coefficients $d=(c_{J_0k},d_{jk},
j=J_0,\ldots,J, k=0,\ldots,\break 2^j-1)$
%
\begin{equation}
p(d | \gamma=1) = N(0,\lambda\Lambda).
\label{eq1}
\end{equation}
In principle $\Lambda$ can be found by explicitly computing the linear
operator of the wavelet decomposition. But from a computational point
of view this is unnecessary and undesirable. Instead
\citet{vannuccicorradi99} show how $\Lambda$ can be derived from $V$
as a bivariate wavelet decomposition of $V$.

\subsection{Shrinkage Through Wavelet Sparsity}
\label{secsparsity}

One of the important advantages of wavelet bases over alternative bases for
$L^2$ functions is the parsimony property of wavelet representations.
Reasonably regular functions are well approximated with only few
nonzero wavelet coefficients. Therefore\break ``shrinkage toward smoothness''
can also be
induced by setting many of
the wavelet coefficients to be zero. We thus assume positive prior
probability for vanishing wavelet coefficients.

Let $\gamma=(\gamma_1,\ldots,\gamma_l)$
denote the vector of indices of nonzero wavelet coefficients,
that is, $d_{jk}=0$ iff $(jk) \notin\gamma$.
We define a prior distribution on $\gamma$ with\vadjust{\goodbreak} geometrically
decreasing probability for nonzero wave\-let coefficients in higher
levels of detail $j$:
\[
\operatorname{Pr}(d_{jk}=0) = 1-\alpha^{j+1}.
\]
See, for example, \citet{abramovichal98} for a discussion of the
choice of $\alpha$.

We write
$\theta_\gamma$ for the subvector of nonzero wavelet coefficients
$d_{jk}$,
and we use $\gamma=1$ for the full model which includes all
coefficients $\gamma=((jk), j=J_0,\ldots,J$  and $k=0,\ldots,2^j-1)$.
The prior $p(\theta_\gamma\vert\gamma)$ for the wave\-let
coefficients under
model $\gamma$ is implied from~(\ref{eq1}) by conditioning the
multivariate normal on $\theta_h=0$, \mbox{$h \notin\gamma$}.
Let $\Omega=V^{-1}$ and write $\Omega_{(\gamma)}$ for the submatrix
with rows and columns $(\gamma_1,\ldots,\gamma_l)$.
Then
%
\begin{equation}
p(\theta_\gamma\vert \gamma) = N\bigl(0,\lambda\Omega_{(\gamma
)}^{-1}\bigr) =
N(0,\lambda\Lambda) .
\label{eqpr-g}
\end{equation}
We use $\Lambda$ to generically denote $\Omega_{(\gamma)}^{-1}$,
suppressing the dependence on $\gamma$ to simplify notation.

\subsection{Illustration of the Shrinkage Effects}

\begin{figure*}

\includegraphics{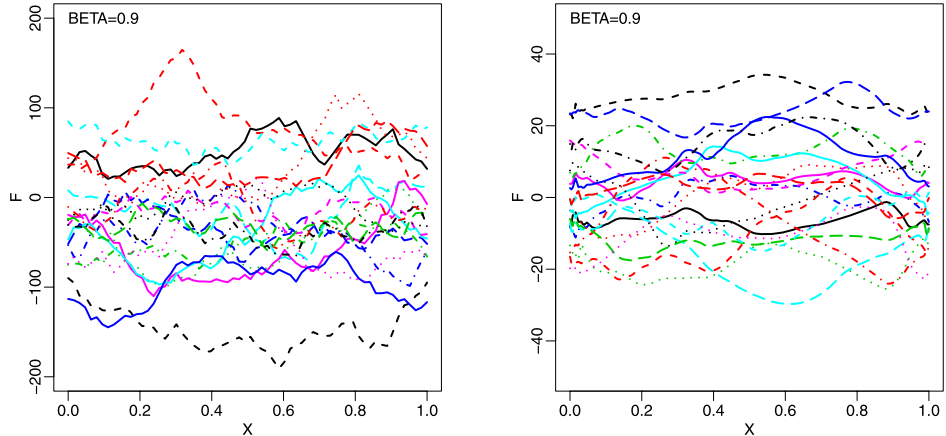}

\caption{
Prior simulations as in Figure
\protect\ref{figprior}, but using $\beta=0.9$ (very little dependence).}
\label{figprior9}
\end{figure*}

Figures \ref{figprior} and \ref{figprior9} demonstrate the ``shrinkage
toward smoothness'' behavior of the priors in Sections~\ref{secsubspace}
and~\ref{secsparsity}. The figures give realizations from the priors
specified in
the two subsections.
Figure~\ref{figprior} utilizes $\beta=0.1$ from the prior in Section~\ref{secsubspace}
and Figure~\ref{figprior9} utilizes $\beta=0.9$. The smaller $\beta$
induces much more
dependence, clearly resulting in smoother functions.

%

The left panel of each figure is generated from use of only the prior
in Section \ref{secsubspace},
that is, all the wavelet coefficients are kept. In contrast, the right
panels of each figure show
what happens when many of the wavelet coefficients are set to zero.
(For simplicity, these were
produced using a standard wavelet thresholding rule.) Clearly, setting
many wavelet
coefficients to zero does seem to result in considerable additional
shrinkage toward smoothness.

%
\vspace*{3pt}\section{Posterior Simulation}\vspace*{3pt}

We implement posterior inference using Markov chain Monte Carlo
simulation.
Marginalizing over~$\theta_\gamma$, we use the posterior
probabilities $p(\gamma\vert y)$ to define a~Metropolis--Hastings scheme
which proposes moves in the model space by adding or deleting
one wavelet basis function at a time.
The computational effort of the proposed scheme is comparable to that of
\citet{georgemcculloch97} and
\citet{smithkohn96}, who suggest schemes based
on algorithms by
\citet{chambers71}
and
(\citeyear{dongarraal79})
which allow fast updating of a Choleski
decomposition of the cross-product matrix $X'X$.
The algorithms proposed by
\citet{georgemcculloch97}
and
\citet{smithkohn96}
allow computation of marginal posterior probabilities with $O(q^2)$
basic operations, whe\-re~$q$ is the number of covariates (basis
functions) included in the model.
We describe a similar efficient updating algorithm in a form suitable
for the wavelet regression problem.\looseness=1
%

\textit{Notation}.
Let $A_{ij}$ be the element in the $i$th row and $j$th
column of a matrix $A$, with $A_i$ being its $i$th column vector.
For a vector $\gamma=(\gamma_1,\ldots,\gamma_l)$ we denote with
$A_\gamma$ the submatrix consisting of \textit{columns}
$(\gamma_1,\ldots,\gamma_l)$, with $A_{(\gamma)}$ the submatrix
consisting of \textit{columns and rows} $(\gamma_1,\ldots,\gamma
_l)$, and
with $A_{(-\gamma)}$ the submatrix with rows and columns
$\gamma=(\gamma_1,\ldots,\gamma_l)$ removed.

Let $x_i,y_i$, $i=1,\ldots,N$, denote the observed data.
Let $h=1,\ldots,2^J$ index the wavelet coefficients
$d=(c_{J_0k},d_{jk})$ and
let $X$ denote the design matrix
\[
X_{ih} =
\cases{
\psi_{jk}(x_i) & $\mbox{for } h=2^{J_0}+1,\ldots,n,$\vspace*{3pt}\cr
\phi_{J_0k}(x_i) & $\mbox{for } h=1,\ldots,2^{J_0} ,$}
\]
where $(jk)$ are the wavelet indices corresponding to the $h$th
element in the vector $d$ of wavelet coefficients.

\textit{Likelihood.}
For a given model $\gamma$ the wavelet decomposition of the unknown
velocity curve $f$ implies a likelihood
%
\begin{equation}
y_i \vert \theta,\gamma\iid N(X_\gamma\theta_\gamma, S),\quad
i=1,\ldots,N ,
\label{eql}
\end{equation}
where $S = \operatorname{diag}(\sigma_i^2)$ with known variances $\sigma^2_i$,
$i=1,\ldots,N$.

\textit{Posterior}.
Together with prior (\ref{eqpr-g}) the likelihood implies a
multivariate normal posterior
$p(\theta_\gamma\vert y,\gamma) = N(\mu,\Sigma)$ with
\begin{eqnarray*}
\Sigma^{-1}& =&
\underbrace{(X_\gamma)'S^{-1}X_\gamma}_{Q^\gamma} +
1/\lambda \Omega_{(\gamma)}
\quad\mbox{and}\\[3pt]
\mu&=& \Sigma\cdot\underbrace{(X^\gamma)'S^{-1}y}_{v^\gamma}.
\end{eqnarray*}
Again, to simplify notation we suppress the dependence on $\gamma$ in
$\mu$ and $\Sigma$.

\subsection{Down Move}\vspace*{1pt}
$\!\!$Assume $\gamma\,{=}\,(\gamma_1,\ldots,\gamma_l)$ and consider a move
``down'' to the submodel $\gamma^*=(\gamma_1,\ldots,\gamma_{l-1})$.
Partition $\Sigma$ into
\[
\Sigma=
\left[
\matrix{
\Sigma_{(-l)} & \tS_l \vspace*{2pt}\cr
\tS_l' & \Sigma_{ll}}
\right]
\]
and similarly $\mu=(\mu_{(-l)},\mu_l)$.
Then
\[
p(\theta_{\gamma^*} \vert y,\gamma^*) = N(\mu^*,\Sigma^*) ,
\]
with $\Sigma^* = \Sigma_{(-l)} - \tS_l \Sigma_{ll}^{-1} \tS_l'$
and $\mu^* = \mu_{(-l)} +\break \tS\Sigma_{ll}^{-1}(-\mu_l)$.
Similarly,
$\Lambda^* = \Lambda_{(-l)} - \tL\Lambda_{ll}^{-1}\tL_l'$.

The corresponding ratio of marginal probabilities is
\[
\frac{p(y \vert\gamma^*)}{p(y \vert\gamma)} =
\biggl(\frac{\lambda \Lambda_{ll}}{\Sigma_{ll}}\biggr)^{1/2}
e^{-(1/2) \mu_l^2/\Sigma_{ll}} .
\]
This expression is easily verified using the candidate formula
$p(y \vert\gamma) =
p(\theta_\gamma\vert\gamma) p(y \vert\theta_\gamma,\gamma)/
p(\theta_\gamma\vert y,\gamma)$ and\break substituting $\theta_\gamma=0$.

\begin{figure*}[t!]
\centering
\begin{tabular}{@{}cc@{}}

\includegraphics{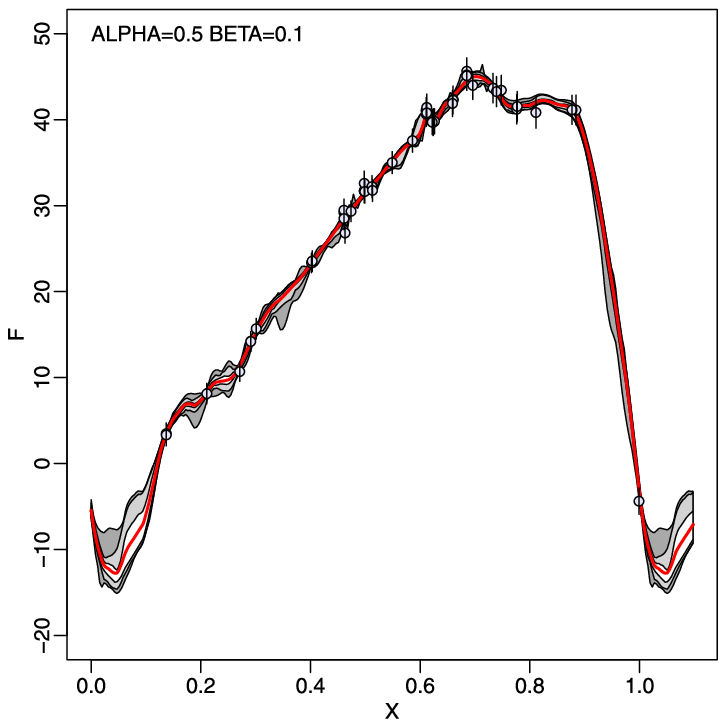}
 & \includegraphics{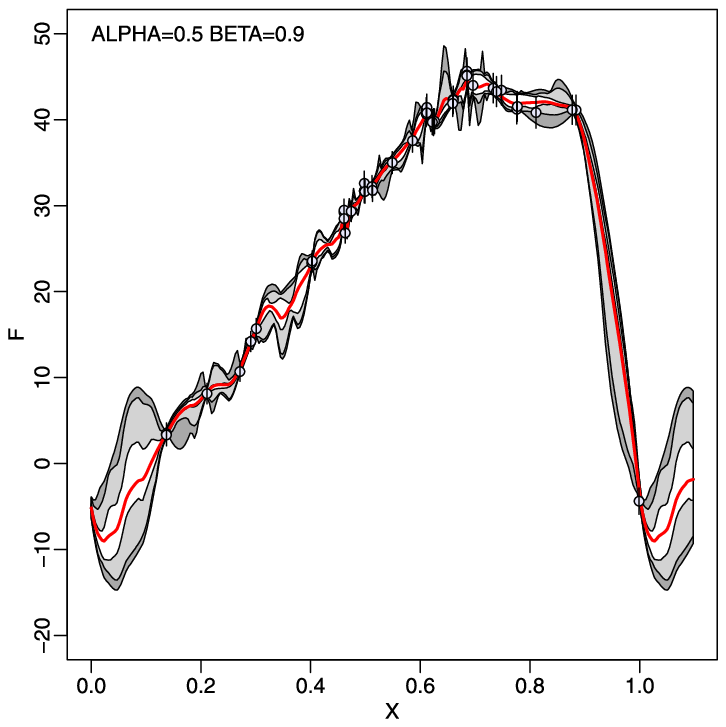}\\
\footnotesize{(a) $\alpha=0.5$, $\beta=0.1$} & \footnotesize{(b) $\alpha=0.5$, $\beta=0.9$}\\[6pt]

\includegraphics{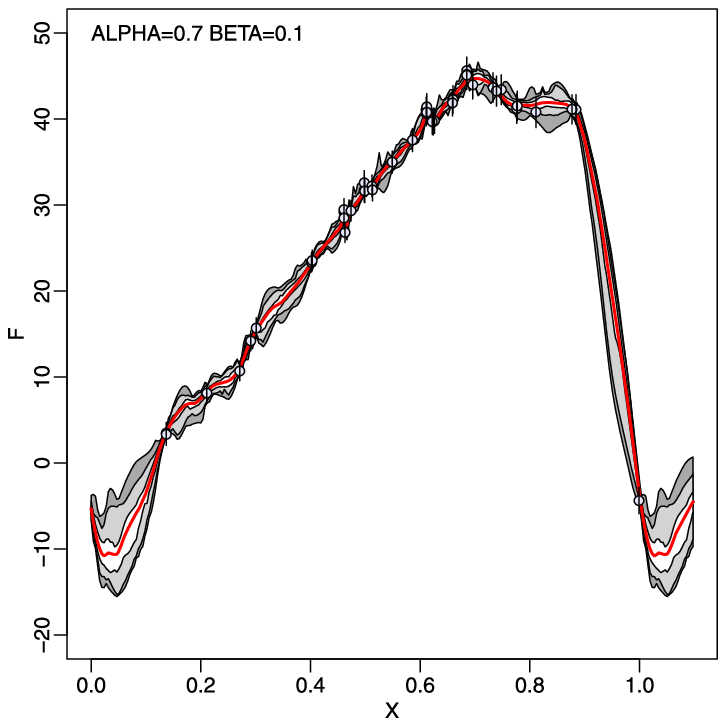}
 & \includegraphics{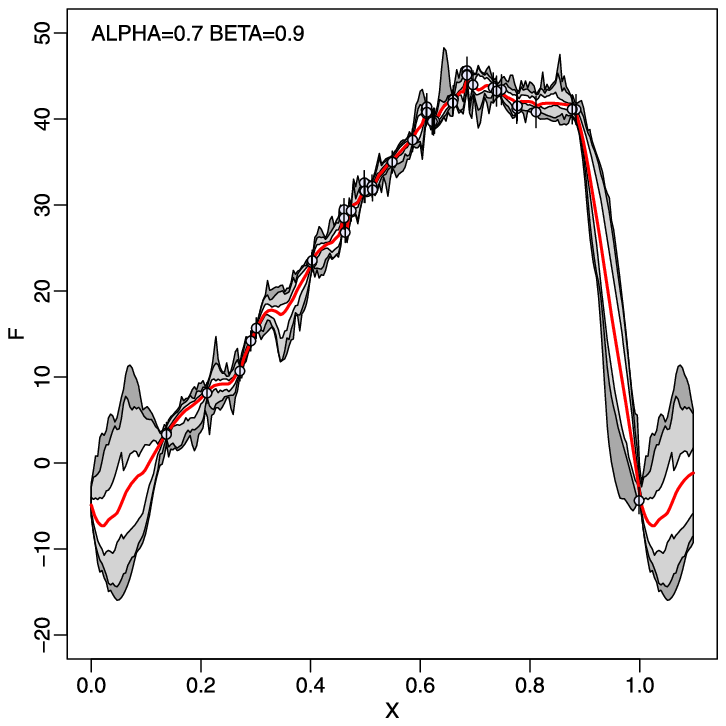}\\
\footnotesize{(c) $\alpha=0.7$, $\beta=0.1$} & \footnotesize{(d) $\alpha=0.7$, $\beta=0.9$}
\end{tabular}
\caption{Posterior inference for T Moncerotis. In all four panels, the thick
smooth line shows the posterior mean
curve. The gray shaded margins show central 50\% (light gray) and central
90\% (dark gray) intervals. The points are the observed data points,
with little error bars showing 2 standard deviations for the
measurement error. Panel \textup{(a)} shows inference under
$\beta=0.1$ and $\alpha=0.5$.
Panels \textup{(b)} through \textup{(d)} show posterior inference using $\beta=0.9$ (\textup{b}
and \textup{d}) and $\alpha=0.7$ (\textup{c} and \textup{d}).
Fixing $\beta=0.9$ essentially assumes independence of the $d_i$ and
implies less
smoothing; setting $\alpha=0.7$ greatly decreases the number of
wavelet coefficients set to zero.}
\label{fig4}
\end{figure*}

\vspace*{1pt}\subsection{Up Move}\vspace*{1pt}
Consider a move from $\gamma$ to $\gamma^*=(\gamma^*_1,\gamma)$.
Denote with $(\mu,\Sigma)$ and $\Lambda$ the posterior and prior
moments under the (current) model $\gamma$:
\[
p(\theta_\gamma\vert \gamma,y) = N(\mu,\Sigma) \quad\mbox{and}\quad
p(\theta_\gamma\vert \gamma) = N(0,\lambda\Lambda).
\]
Similarly, let $(\mu^*,\Sigma^*)$ and $\Lambda^*$ denote the posterior
and prior moments under the (proposed) model $\gamma^*$:
\begin{eqnarray*}
p(\theta_{\gamma^*} \vert \gamma^*,y) &=& N(\mu^*,\Sigma^*)\quad
\mbox{and}\\
p(\theta_{\gamma^*} \vert \gamma^*) &=& N(0,\lambda\Lambda^*).
\end{eqnarray*}
For posterior simulation we use a lower triangular Choleski
decomposition of the posterior variance/\break covariance matrix,
$T T'=\Sigma$ and $T^{*\prime} T^* = \Sigma^*$.
The new moments $\mu^*,\Sigma^*$ and $\Lambda^*$ and the Choleski
decomposition $T^*$ are computed using the
following expressions.

Let
$Q^*=(X^{\gamma^*})'S^{-1}X^{\gamma^*},
\Omega^*=\Omega_{(\gamma^*)},
Q=(X^{\gamma})'\cdot\break S^{-1} X^{\gamma}$ and
$\Omega=\Omega_{(\gamma)}$
and partition
\[
Q^* =
\left[
\matrix{
Q^*_{11} & \tQ^{*\prime}_1 \vspace*{2pt}\cr
\tQ^{*}_1 & Q}
\right]
\quad\mbox{and}\quad
\Omega^* =
\left[
\matrix{
\Omega^*_{11} & \tO_1^{*\prime} \vspace*{2pt}\cr
\tO_1^* & \Omega}
\right].
\]
Let $b=\tQ^*_1 + 1/\lambda\tO_1^*$,
$h=\Sigma b$,
$c=\tQ_{11}^* + 1/\lambda \Omega^*_{11}$,
$b_0=\tO_1^*$,
$h_0 = \Lambda\tO_1^*$ and
$c_0 = \Omega^*_{11}$.
Then
\begin{eqnarray*}
\Sigma^*& =&
\left[
\matrix{
0 & 0\vspace*{2pt}\cr
0 & \Sigma}
\right]
+ \frac1{c-b'h}
\left[
\matrix{
1 & -h'\vspace*{2pt}\cr
-h & h h'}
\right]
\quad\mbox{and}
\end{eqnarray*}
\begin{eqnarray*}
\hspace{-8pt}
\Lambda^* &=&
\left[
\matrix{
0 & 0\vspace*{2pt}\cr
0 & \Lambda}\right]
+ \frac1{c_0-b_0'h_0}
\left[
\matrix{
1 & -h_0'\vspace*{2pt}\cr
-h_0 & h_0 h_0'}
\right],
\\[3pt]
\hspace{-8pt}
\mu^* &=&
\pmatrix{
0\vspace*{2pt}\cr \mu}
 +
(c-b'h) \Sigma^*_1 \Sigma^{*\prime}_1 v^{(\gamma^*)},
\end{eqnarray*}
and
$T^*$ is obtained by augmenting $T$ with a new first column
$w=\Sigma_1^*/\sqrt{\Sigma_{11}^*}$ to
\[
T^* =
\left[
\matrix{
& 0\vspace*{2pt}\cr
w & T
}
\right].
\]
%
The corresponding ratio of marginal probabilities is, by symmetry to
the down move,
\[
\frac{p(y \vert\gamma)}{p(y \vert\gamma^*)} =
\biggl( \frac{\lambda\Lambda^*_{11}}{\Sigma^*_{11}}\biggr)^{1/2}
e^{-(1/2) \mu_1^{*2}/\Sigma^*_{11}} .
\]


\section{Example}

We apply the above methodology to the data for the star T Moncerotis,
as shown
in Figure \ref{TMon}, for the choices $\beta=0.1$ (strong dependence of the $d_i$)
and $\alpha=0.5$ (inducing a moderate level of sparsity). The resulting
nonparametric posterior is difficult
to summarize; some features of this posterior are presented in Figure
\ref{fig4}(a).

It is, of course, one of the strengths of the Bayesian approach to shrinkage
that uncertainty in the shrinkage estimate [the posterior mean of
$f(x)$, given by the
thick center line in Figure \ref{fig4}(a)] can also be given. This is
crucial in
characterizing the (considerable) uncertainty in the eventual estimate of
distance to the star (see \citep{barnesal03}).

Figure \ref{fig4} also indicates the effect on the T Moncerotis data
of each of the shrinkage priors in Sections \ref{secsubspace} and~\ref{secsparsity}.
Panel (b) shows the effect of the prior
in Section \ref{secsubspace}; setting $\beta=0.9$ effectively
makes the $d_i$ independent. Panel (c) shows the effect of the
prior in Section \ref{secsparsity}; setting $\alpha=0.7$ greatly decreases the
number of
wavelet coefficients set to zero. In both cases, the
posterior functions appear to be unreasonably rough
and the uncertainty in the shrinkage estimate appears
to be unreasonably large. Pa\-nel~(d), which effectively uses neither of the
shrinkage techniques, is especially unsatisfactory.

\section*{Acknowledgments}
This research was supported in part by NSF Grants DMS-01-03265,
DMS-06-35449 and
DMS-07-57549-001. We are grateful to Thomas Barnes for providing us
with the data analyzed herein.

%

\end{document}